\documentclass[graybox, envcountchap, authoryear, natbib]{svmult}

\usepackage{mathptmx}       
\usepackage{helvet}         
\usepackage{courier}        
\usepackage{type1cm}        
\usepackage{makeidx}         
\usepackage{graphicx}        
\usepackage{multicol}        
\usepackage[bottom]{footmisc}

\newcommand{\hi}{H\,I}
\newcommand{\kms}{km~s$^{-1}\:$}
\newcommand{\arcdeg}{^{\circ} }

\newcommand{\Msol}{M$_{\odot}$} 

\newcommand{\cms}{cm$^{-2}$}
\newcommand{\mhi}{${\rm M_{HI}}$}
\newcommand{\Nhi}{${\rm N_{HI}}$}

\begin{document}

\title*{Neutral Gas Accretion onto Nearby Galaxies}
\author{Felix J. Lockman}
\institute{Felix J. Lockman \at Green Bank Observatory, \email{jlockman@nrao.edu}. The Green Bank Observatory is a facility of the National Science Foundation, operated under a cooperative agreement by Associated Universities, Inc.}

%
%
\maketitle

\abstract*{While there is no lack of evidence for the accretion of stellar systems onto nearby galaxies, direct evidence for the 
accretion of gas without stars is scarce.  Here we consider an inventory of starless gas ``clouds" in and around galaxies of the 
Local Group to discern their general properties and see how they might appear in distant systems.  The conclusion is that 
accreting gas without stars is detected almost entirely within the circumgalactic medium of large galaxies and is rare otherwise.  
If our Local Group is any example, the best place to detect starless gas clouds is relatively close to galaxies. 
}

\abstract{While there is no lack of evidence for the accretion of stellar systems onto nearby galaxies, direct evidence for the 
accretion of gas without stars is scarce.  Here we consider an inventory of starless gas ``clouds" in and around galaxies of the 
Local Group to discern their general properties and see how they might appear in distant systems.  The conclusion is that accreting
gas without stars is detected almost entirely within the circumgalactic medium of large galaxies and is rare otherwise.  
If our Local Group is any example, the best place to detect starless gas clouds is relatively close to galaxies. 
}

\section{Galaxies Then and Now}
\label{sec:Introduction}

\begin{quotation}
{\it The evidence at present available points strongly to the conclusion that the spirals are individual galaxies, or island universes, comparable with our own galaxy in dimensions and in number of component units.}  -- H.D. Curtis
\end{quotation}

\begin{quotation}
{\it ... the extraplanar gas seems to consist of two parts: a large one from galactic fountains and a smaller part accreted from intergalactic space. There is direct (HVCs in our galaxy and filaments in external galaxies) and indirect (rotational velocity gradients) evidence for the accretion from outside.} --- Sancisi et al. 2008
\end{quotation}

\begin{quotation}
{\it Galaxies are like people.  Every time you get to know one well, it turns out to be a little peculiar.} -- Sidney van den Bergh
\end{quotation}

\begin{figure}[!b]
\includegraphics[width=4.5in]{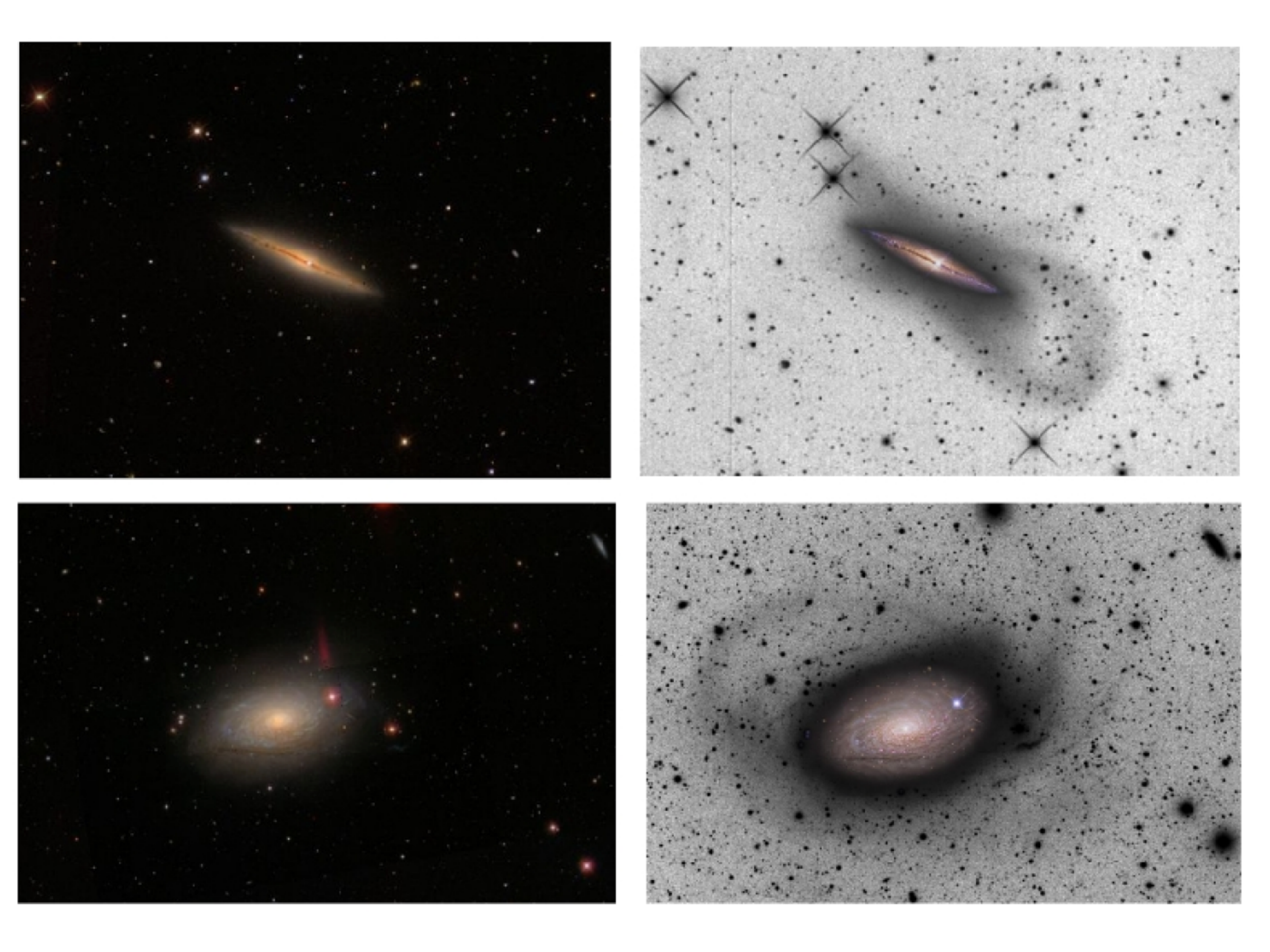}
\caption{Examples of stellar accretion in two galaxies from \cite{Carlin2016}.   Images on the left from the 
Sloan Digital Sky Survey of the nearby galaxies NGC 4013 (top) and M 63 (bottom) show regular disks while the much deeper images on the right reveal streams of stars accreted from smaller galaxies.
}
\label{fig:Carlin}       
\end{figure}

Unlike the days of the {\it Island Universe}, when galaxies floated in solitary splendor on Hubble's Tuning Fork \citep{Hubble},
   today's galaxies are a mess (Fig.~\ref{fig:Carlin}).  
Evidence for the growth and evolution of galaxies by the capture of stellar systems is everywhere and there are 
arguments for continued accretion of gas as well.  Certainly some gas will arrive with the 
small stellar systems that large galaxies devour, but how does it get to the disk, and is there neutral gas accreting from other sources?  
In this article I consider the evidence for accretion of neutral gas 
 onto nearby galaxies, especially gas that is not associated with stars.  
 This is not a comprehensive review, but instead explores the connection between 
 gas likely to be accreting onto galaxies in the Local Group, where we can examine it with high sensitivity and 
 linear resolution, and that seen in more distant systems with a better vantage 
 but considerably less sensitivity.  Is there local accretion of starless gas, and what form does it take?

Two comprehensive reviews are directly relevant to this topic.  
Sancisi et al. \cite{Sancisi2008} consider the 
evidence for accretion of neutral gas  by analyzing high resolution  \hi\ maps of several dozen nearby galaxies.  They find ample 
evidence of kinematic anomalies, tails and filaments, warps, lopsided disks, and interaction.  They adopt the stance that any 
significant deviation of a galaxy's \hi\ from symmetry is evidence for interaction, and that interaction implies accretion.
Accretion also has a prominent place in the recent review of gaseous galaxy halos by \cite{Putman2012} which takes a thorough look at circumgalactic gas in all phases.   
I will refer to these reviews throughout this article for more complete discussion of some topics.

The volume around galactic disks contains material in many forms.  Both the Milky Way (MW) and M31 have a 
hot $10^{6-7}$ K circumgalactic medium (CGM) of enormous mass and extent, possibly dwarfing the baryon content of the disks themselves \citep{AndersonBregman,Hodges-Kluck2016,Lehner2015}. 
This gas may cool and condense, feeding the disk.  There may also be cosmological accretion -- cold flows --  
which may have to traverse the CGM to reach the disk and in that passage may be disrupted entirely \citep{Putman2012}.  
Then there is gas 
that has been stripped from one galaxy through interaction and ultimately ends up in the disk of another.  Add to that the ejection of 
gas from a galaxy's disk from supernovae or a nuclear wind, and the processes can  become quite difficult to disentangle.  
Considerable insight into the the CGM of galaxies has come through studies of Ly$\alpha$ and MgII absorption lines
\citep{Kacprzak2013,Wakker2009}, but those data will not be discussed here.  Instead, 
we will concentrate on 21cm \hi\ observations, and work from the inside out, 
from the interstellar medium (ISM)  disk-halo interface, to high velocity clouds (HVCs), to the products of interaction between 
galaxies, and finally to \hi\  "clouds" that don't easily fit into any of these categories.  

But first a note on the detectability of \hi.  Filled-aperture (single dish) radio telescopes can easily detect \hi\ column densities of 
${\rm N_{HI} = 10^{18}}$ \cms, and with a little work, $10^{17}$ \cms.  This comes at the expense of angular resolution, which for modern
instruments is in the range $3^{\prime} - 10^{\prime}$.  Hydrogen clouds with ${\rm M_{HI} \geq 10^5}$ \Msol\ can thus be detected anywhere in the Local 
Group at a linear resolution of 1--3 pc d$_{\rm kpc}$.   Aperture synthesis instruments provide much higher angular resolution of $ <1^{\prime}$
but at the cost of reduced sensitivity.  
The smallest object in the table of plumes, wings, and other peculiar \hi\ structures of \cite{Sancisi2008}  has  $10^8$ \Msol.
The deepest aperture synthesis \hi\ observations of galaxies so far have come from the HALOGAS survey 
\cite{Heald2011} which reaches a limiting sensitivity of ${\rm N_{HI} = 10^{18.5-19.0}}$ \cms.  

This sensitivity gap is not unbridgeable, but we should keep it in mind when comparing local with distant objects.  

\section{The Disk-Halo Interface: Clouds and Shells }
\label{sec:Disk-Halo}

\begin{figure}[b]
\includegraphics[width=4.5in, clip=true]{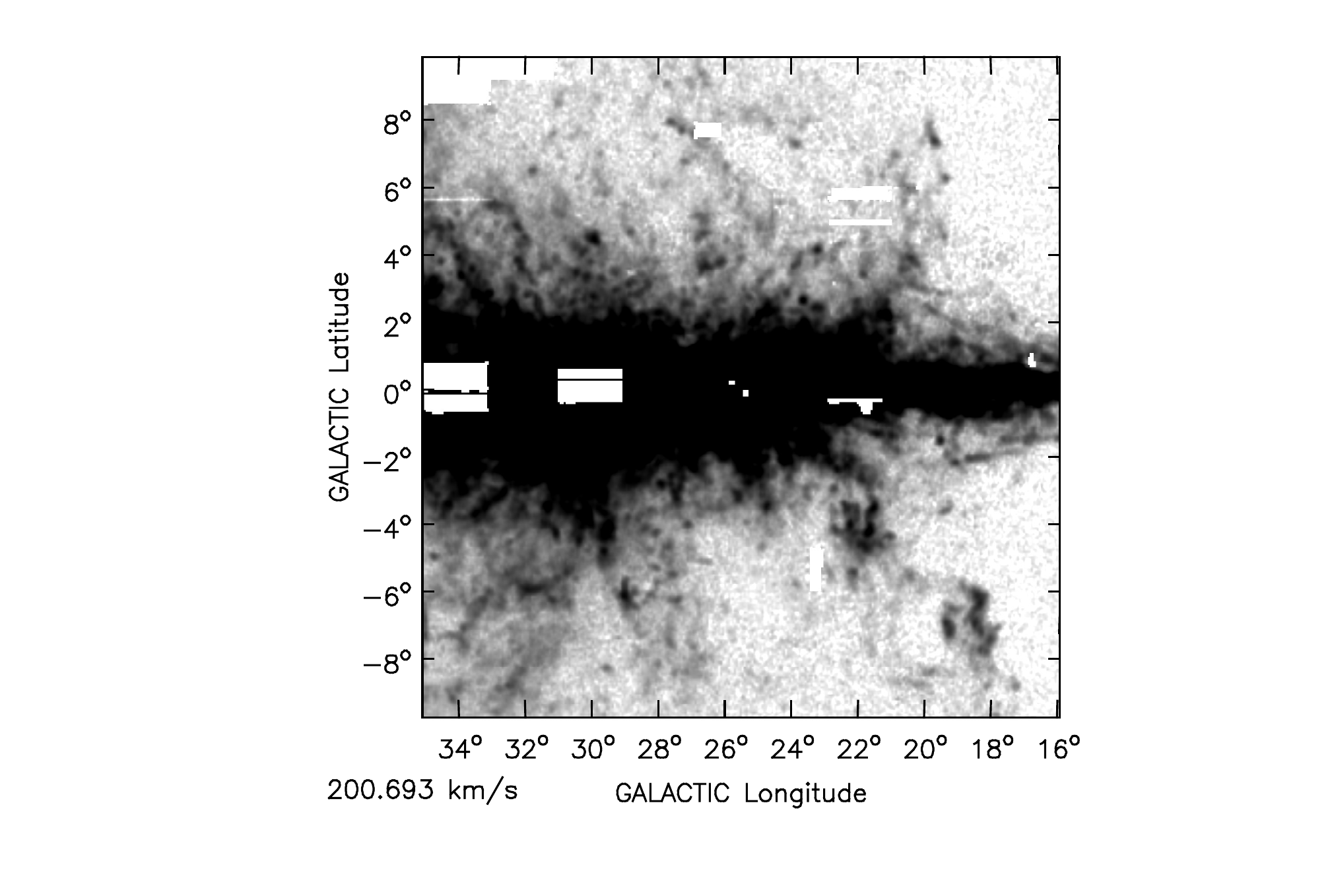}
\caption{Cut through the Galactic plane at velocities approximately along the tangent point showing the vertical 
extent of Galactic HI.  In this projection a latitude of  $8\arcdeg$ corresponds to a distance $\sim 1$ kpc from the Galactic plane. 
These observations can resolve structures down to a few 10s of pc in size.}
\label{fig:2}       
\end{figure}

The Milky Way gives us a close view of process like the galactic fountain, which lifts gas several kpc away from a spiral galaxy's disk 
\citep{Shapiro1976,Bregman1980}.
Figure~\ref{fig:2} shows Milky Way \hi\ along a cut $\sim 20\arcdeg$ in longitude through the 
Galactic plane approximately along the tangent points.   It is easy to find \hi\ loops, filaments, clouds and a diffuse 
component extending many degrees away from the plane.  In this figure 
each degree of latitude corresponds to a displacement  $\approx 135$ pc from the midplane, so 
the \hi\ emission at $b = \pm 4\arcdeg$ arises in neutral gas at a height $z \sim 0.5$ kpc,  several \hi\ scale heights 
above the main \hi\ layer \citep{Dickey1990}.
  In the inner Milky Way about $10\%$ of the \hi\ can be found  more than 0.5 kpc from the disk \citep{Lockman1984}.
 Some of this \hi\  is contained  in  discrete clouds that can be identified to $z \approx 2$ kpc 
  \citep{Lockman2002,Ford2010}. Some of the extraplanar \hi\ is also organized  
as large "supershells", whose tops can reach $|z|>3$ kpc.  They can be $\sim 1$ kpc is size and contain $3 \times 10^4$ \Msol\ of  neutral gas 
\citep{Heiles1979,McClure-Griffiths2002,Pidopryhora2007}.  
 
Supershells certainly have their origin in supernovae and stellar winds  \citep{Tomisaka1986}, 
and can often be linked directly to sites of star formation \citep{Pidopryhora2007,Kaltcheva2014}.  
The cloud-like component of extraplanar \hi\ has a less certain origin, though it may result from the breakup of shells.
Often referred to as "disk-halo" clouds, they appear to be related both spatially and kinematically 
to the disk, though the population extends well into the lower halo \citep{Lockman2002,Ford2010}. 
Some of the larger Milky Way disk-halo clouds  with ${\rm M_{HI} \approx 500}$ \Msol\ 
have been imaged at 2-3 pc resolution and show a variety of shapes with 
some sharp boundaries and evidence of a two-component thermodynamic structure \citep{Pidopryhora2015}.  
Note that from the above discussion on sensitivity, these disk-halo clouds could not be detected beyond the Milky Way 
as individual objects.  
Within the Milky Way there is a general correlation between the amount and vertical extent of disk-halo clouds and the spiral 
arms \citep{Ford2010} although there is no detailed correlation with individual star-forming regions.   

A similar extraplanar \hi\ layer is seen in a number of galaxies, often containing 10\% of the disk \hi\ mass, though 
the percentage has large variations from galaxy to galaxy \citep{Sancisi2008}.
While it is not possible to isolate individual disk-halo clouds in other galaxies, 
the better vantage afforded in observations of extragalactic systems allows the kinematics of this component to be 
analyzed more completely.  The disk-halo (or extraplanar) gas  often shows evidence of a vertical lag in 
rotational velocity of $10-20$ km s$^{-1}$ kpc$^{-1}$ but with a large range; 
sometimes an inflow toward the center of a galaxy is also inferred at the level of 
10-20 \kms\   \citep{Fraternali2002,Sancisi2008,Zschaechner2015}.  
This extraplanar gas does not show large deviations from prograde galactic rotation, 
 and constitutes a dynamic neutral galactic atmosphere.  
 
 The disk-halo 
gas extends to heights where it may mix with material accreting through or cooling from the CGM \citep{Putman2012}.  
Thus the presence of neutral gas many scale-heights away from the plane 
is not unexpected, and distinguishing between gas that is "recycled" 
as much of the disk-halo material must be, and gas that has never been in the disk,  
may not be straightforward in the absence of other information.  Although the disk-halo clouds 
are concentrated to the disk in the sense that their numbers increase towards the disk, 
there is no way to know if any individual cloud had its origin in the Milky Way, or has been 
accreted, or is a combination of both processes \citep{Marasco2012}. 
  It would be very interesting to have information on the elemental abundances in disk-halo clouds at different heights.

Any gas accreting onto the Milky Way has to pass through the extended disk-halo layer before it reaches the inner disk.
 One massive cloud passing through this layer shows evidence of disruption, and is discussed in section \ref{sec:Smith}.

\section{High Velocity Clouds}
\label{sec:HVC}

It has been known for more than 50 years that there are significant amounts of neutral hydrogen
 around the Milky Way that do not follow
Galactic rotation -- the high velocity clouds (HVCs) \citep{Muller1963,Wakker2004}.  
For many years the lack of information on their distances has allowed speculation that some of them might be 
  $\sim1$ Mpc or more away from the Milky Way and thus have ${\rm M_{HI} = 10^{7-8}}$ \Msol\ 
  \citep{Blitz1999,Braun1999}.  However,  the discovery of a similar population around  
M31, M33, and other galaxies \citep{Thilker2004,Westmeier2008,Grossi2008, Putman2009,Keenan2016,Miller2009}, 
   suggests that the phenomenon is likely 
common and, as importantly, confined to the inner CGM of spirals.    
This does not mean that some of the more compact, 
more isolated HVCs might not be free-floating  independent systems, but the total 
\hi\ mass within the CGM of normal galaxies (aside from the products of tidal  interaction) 
must be relatively small \citep{Pisano2007,Pisano2011}.   The HVC system of the Milky Way 
has been estimated to contain an \hi\ mass of  $\sim 3 \times 10^7$ \Msol\ \citep{Putman2012}. 
The HVC population likely contains as much ionized gas as neutral gas 
\citep{Shull2009,Shull2011,Lehner2011}.

One of the most important facts about HVCs is that 
they do not contain stars \citep{Hopp2007}.   In this 
sense, compact HVCs that are discovered to be dwarf 
galaxies \citep[e.g. Leo P;][]{Rhode2013}, are simply objects that have been  mis-classified as HVCs.   
Another important property is that their velocities are not really that high: none have velocities near or 
greater than the escape velocity of the Milky Way \citep{Wakker2004}. \cite{Putman2012} 
estimate that the larger Milky Way HVCs have kinematics that can be fit by a prograde rotational component 
about one-third the circular velocity of the Sun, and an inflow component of a few tens of \kms, though one 
prominent cloud almost certainly has a large retrograde motion \citep{Lockman2003}.

\subsection{High Velocity Clouds in M31 and M33}

\begin{figure}[b]
\includegraphics[width=4.5in, clip=true]{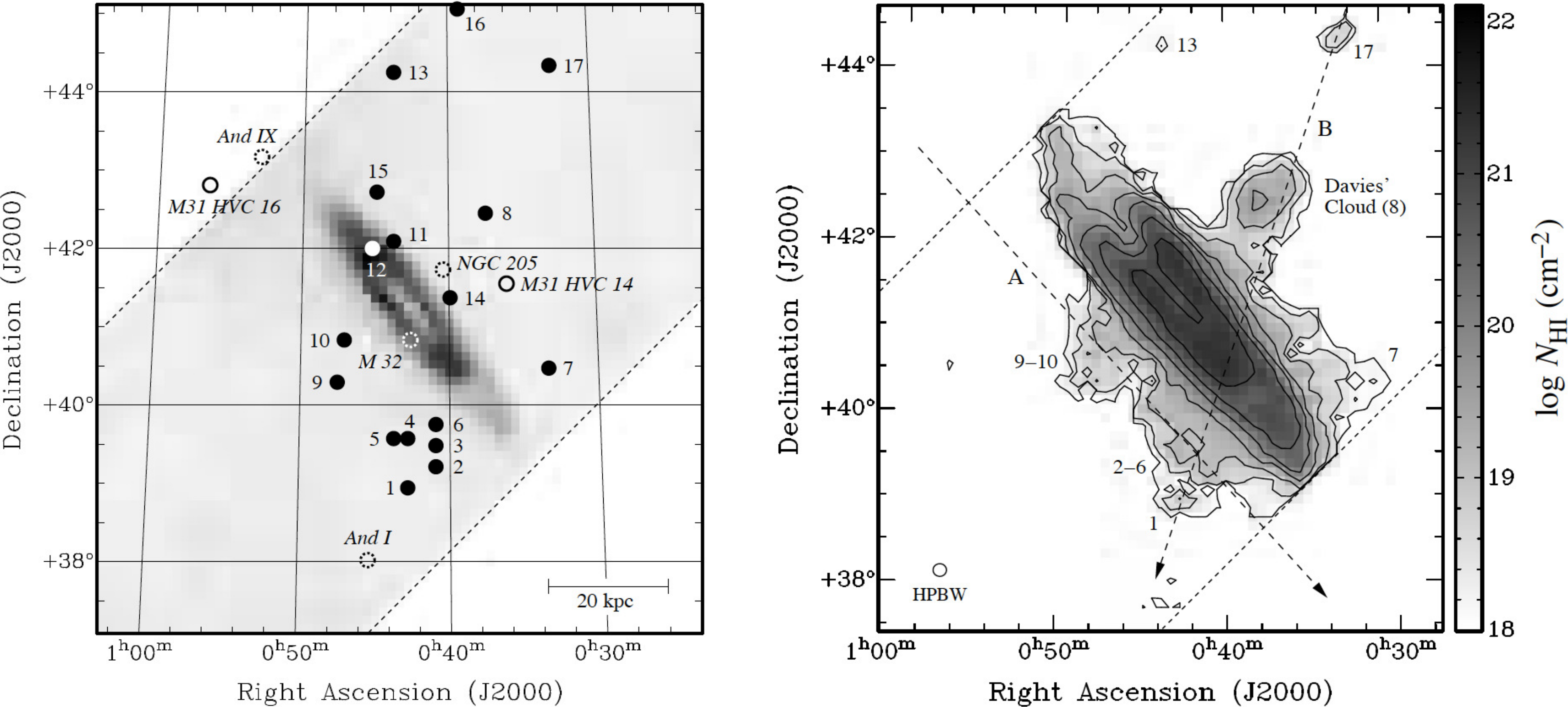}
\caption{From \cite{Westmeier2008}.  Left panel: the HVC system of  M31 in relationship to the bright 
\hi\ in the disk.  Right panel: integrated \hi\ over all velocities with contours at log$_{10}(N_{HI})$ 18, 18.5, 19, 19.5, ... .
Many of the HVCs seem to blend into the disk, but in maps at different 
velocities it is clear that they are distinct objects.
This is how M31 would appear to current instruments were it $\sim 10$ Mpc distant, except that the two lowest 
contours would be missing.  As noted by \cite{Putman2012} this image of M31 bears a strong 
resemblance to the \hi\ image of the galaxy NGC 891 \citep{Oosterloo2007}  which has been observed with about the  
similar linear resolution as M31.
}
\label{fig:Westmeier}       
\end{figure}

The HVC population of M31 is shown in Figure~\ref{fig:Westmeier}.     
The typical M31 HVC  has \mhi\  of a few 10$^5$ \Msol, and would not be detectable as an individual object in 
most measurements of more distant galaxies. 
The right panel of Fig.~\ref{fig:Westmeier} emphasizes the proximity of  
HVCs to that galaxy's disk, 
 and the difficulty one would have in disentangling them from the disk-halo layer in more distant systems.  
The HVCs of M31 seem to be confined to within 50 kpc of that galaxy, though there are indications that some may be found 
to greater distances to the north \citep{Westmeier2008,Wolfe2016}.    The total \hi\ mass in the M31 HVCs is $2 \times 10^7$ \Msol; when 
extraplanar gas possibly associated with the disk is included, the amount may be as high as $5 \times 10^7$ \Msol\ \citep{Westmeier2008}.

There are also HVCs around M33, though here confusion with that galaxy's disk and possibly unrelated extraplanar 
gas makes it difficult to separate the populations cleanly \citep{Grossi2008,Putman2009,Keenan2016}.   
The total \hi\ mass in the M33 HVCs is $3.5 \times 10^7$ \Msol\ using only the data from the most recent study
 \citep{Keenan2016}.  If we include  clouds that may be located in the disk-halo region \citep{Grossi2008,Putman2009}
  the total M33 HVC \hi\ mass increases to $\sim 5 \times 10^7$ \Msol.

\begin{figure}[b]
\includegraphics[width=4.5in]{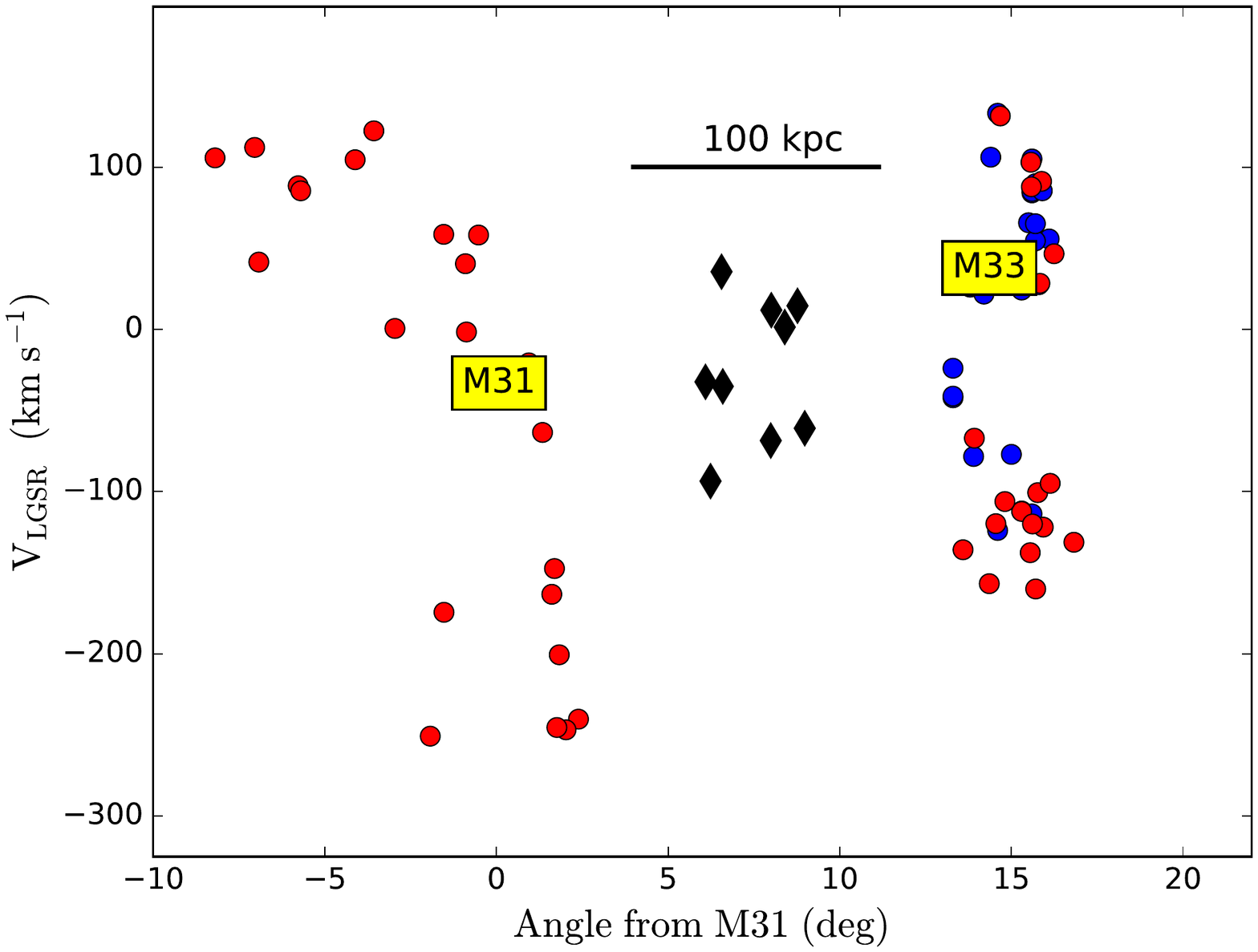}
\caption{The HVC systems of M31 and M33 (red circles), where the velocity with respect to the Local 
Group Standard of rest, ${\rm V_{LGSR}}$, is plotted against the
angular distance from M31 
in the direction of M33.  To the samples of 
M31 and M33 HVCs  \citep{Westmeier2008,Keenan2016}  are added clouds that may be in the halo of 
M33 (blue circles) \citep{Grossi2008,Putman2009}.  It can be difficult to distinguish between the two groups, which may 
actually blend together.  The systemic velocities of M31 and M33 are indicated.  Diamonds mark the M31-M33 clouds 
discussed in section \ref{sec:M31-M33_clouds}.  They have velocities more in common with the systemic velocity of M31 and M33 than 
of their system of HVCs.  There are no stars in any of these clouds.
}
\label{fig:VLGSR}       
\end{figure}

Figure \ref{fig:VLGSR} shows the velocity and location of the HVCs of M31 and M33 with respect to M31.  At the distance of these galaxies 
one degree on the sky corresponds to about 15 kpc.   This figure includes some objects in M33 (marked in blue)
that are probably part of the disk-halo 
interface and not true HVCs \citep{Putman2009,Keenan2016}, but are included to show that aspects 
of these two populations can blend and they often occupy very similar areas of position and velocity space.   
This figure shows another interesting aspect of HVCs.  The velocity spread of those associated with M33 is considerably 
smaller than those associated with M31.  If HVCs are condensations from the CGM this would follow naturally, as M33 is less 
massive than M31 and so presumably has a CGM that rotates more slowly.

A key fact that has been established through distance measurements to Milky Way HVCs and studies of M31 and M33 is that 
HVCs are not free-floating in the Local Group, but are concentrated around the large spirals.  
If the HVCs are even in approximate pressure equilibrium with a galaxy's CGM they must be orders of magnitude 
more dense than the hot gas around them 
and thus falling toward the disk.  There is some evidence for the interaction of Galactic HVCs with the 
CGM in the form of distortions of the cloud shapes \citep{Bruns2000}, 
and there is rather spectacular evidence for the direct 
accretion of \hi\ onto the Milky Way from one HVC: The Smith Cloud.  

\subsection{The Smith Cloud -- Accretion in Action}
\label{sec:Smith}

\begin{figure}[b]
\begin{center}
\includegraphics[scale=.65]{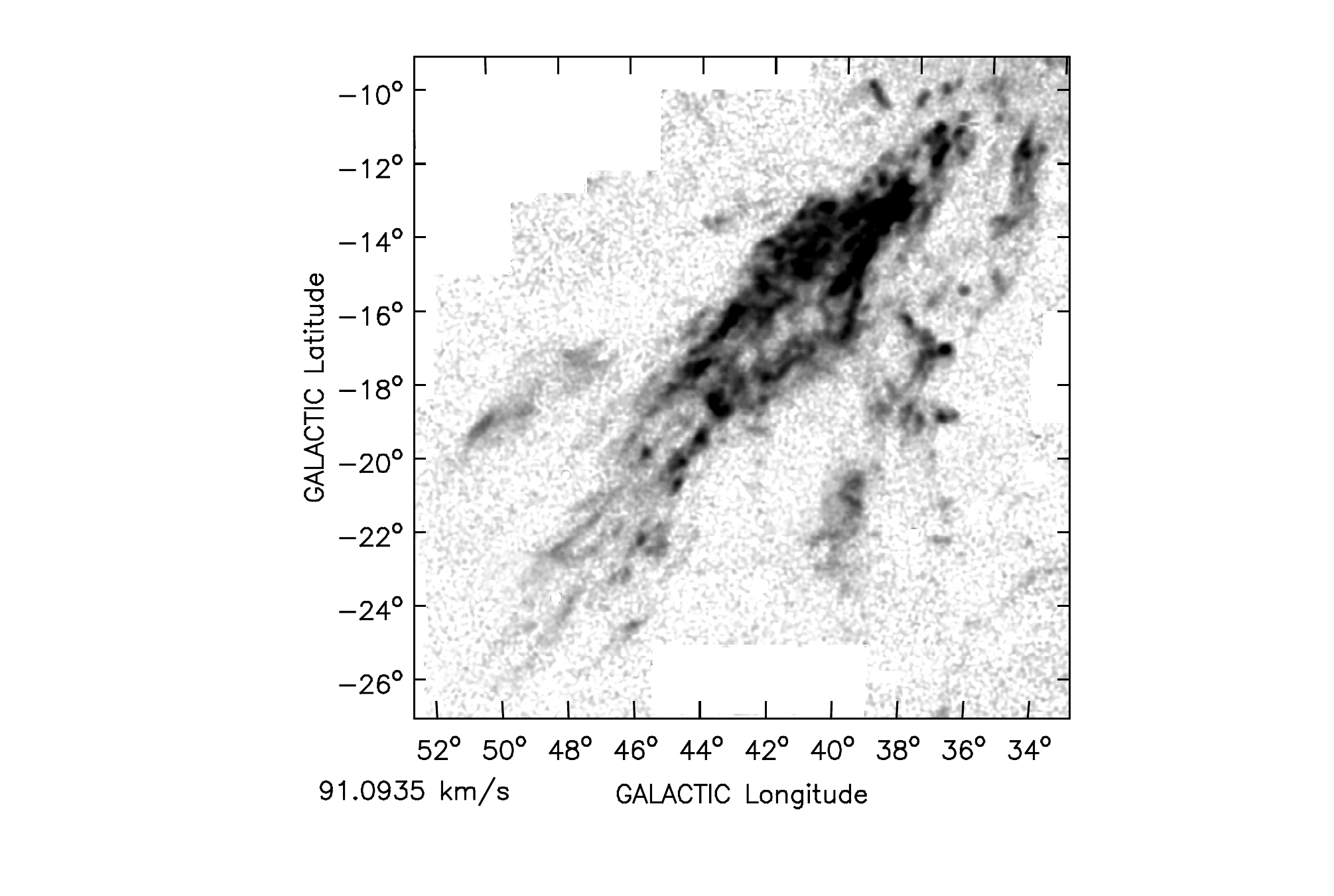}
\end{center}
\caption{Channel map of the Smith Cloud from new GBT data.  All the emission in this  figure is associated 
with the Cloud, which has other components extending down to latitude $-40\arcdeg$.  The part of the Cloud 
displayed in this image is about 4 kpc in length.
The Galactic plane lies  above this figure, and the Cloud is moving toward it at an angle.
}
\label{fig:Smith}       
\end{figure}

In the same year that the discovery of HVCs was announced, 
a short paper appeared reporting observations of
 a peculiar \hi\ feature that now 
appears to be a very important object, the Smith Cloud \citep{Smith1963}.  
Figure~\ref{fig:Smith} shows an \hi\ channel map 
from a recent 21cm survey made with the Green Bank Telescope (GBT).   
The cloud has a good distance estimate \citep{Wakker2008}, 
and thus a well defined mass and size; it is moving toward 
the Galactic plane, which it should intersect in $\sim 30$ Myr if it is survives as a coherent entity \citep{Lockman2008}.  
Its mass in \hi\ is $\sim 2 \times 10^6$ \Msol\ and it has an ionized component with a similar mass 
detected in faint ${\rm H_{\alpha}}$ emission \citep{Hill2009}.
There is evidence that it has a magnetic field $\approx 8 \mu$G \citep{Hill2013} and 
 a S/H abundance that is about one-half Solar \citep{Fox2016}.
Its total space velocity is below the escape velocity of the Milky Way, and 
 the largest component is in the direction of Galactic rotation.  It  
appears to be entering the Milky Way at a rather shallow angle \citep{Lockman2008,Nichols2009}.  
The Smith Cloud is thus adding angular momentum to the disk.  
It has no detectable stars \citep{Stark2015}.  

\begin{figure}[b]
\includegraphics[width=4.5in, clip=true]{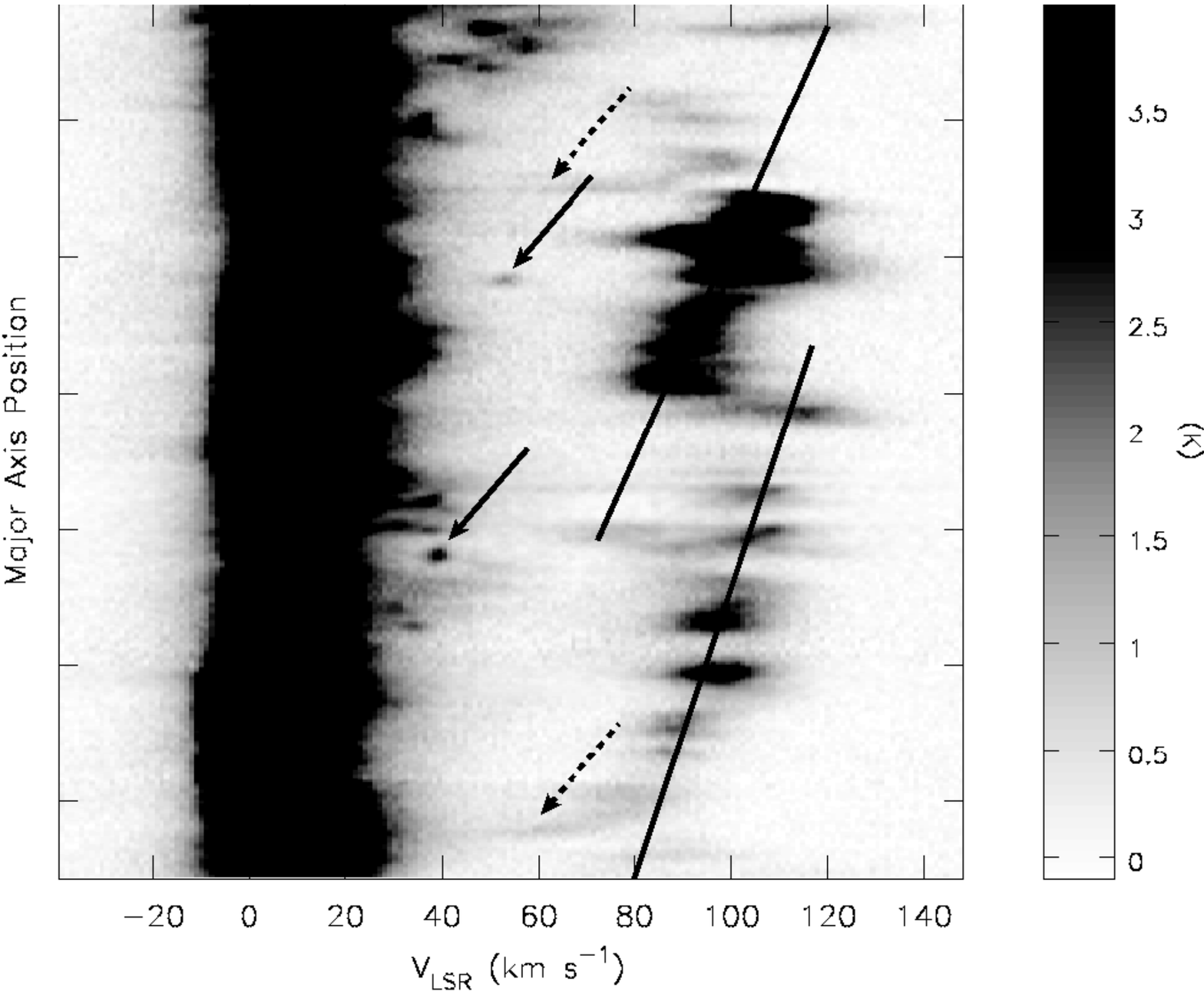}
\caption{A velocity-position plot down the major axis of the Smith Cloud from \cite{Lockman2008}. Arrows mark lumps or streams of 
\hi\ that have been decelerated by interaction with the Milky Way halo.  
Besides a general stripping and deceleration of the cloud edges (\hi\ indicated by dashed arrows), it appears that the Cloud is 
encountering dense lumps in the Milky Way disk-halo interface that remove chunks of the Cloud (noted with solid arrows).  
Some of these correspond in detail to voids in the body of the Smith Cloud. 
}
\label{fig:Smith_vb}       
\end{figure}

The brightest, most compact component of the Cloud lies about 3 kpc below the Galactic plane.  
It is thus in the disk-halo transition, and it appears that it is encountering a clumpy medium.   The Cloud has holes 
  that are matched by small fragments at 60 \kms\ lower velocity, consistent with the 
velocity of the Milky Way's halo at that location (Fig.~\ref{fig:Smith_vb}).  
This object shows clear signs of interaction with the Milky Way's extraplanar gas.

There are many puzzles surrounding the Smith Cloud.  If it has condensed from the CGM than why does it have a 
S/H metallicity ratio higher than typical HVCs \citep{Fox2016,Wakker2004}?   If it originated  from the Galaxy, then how 
did it acquire such a large peculiar motion and mass, which implies a kinetic energy  $\sim 5 \times 10^{53}$ ergs \citep{Marasco2017}?
  Does it require a significant dark matter component to maintain its stability, as suggested by some investigations \citep{Nichols2009,Nichols2014}?

  If the Smith Cloud were at the distance of M31 it would appear to the HVC surveys with 
  a peak \Nhi\  around  $2 \times 10^{19}$ \cms,  a few times higher than the M31 HVCs, but not especially anomalous.  
With an \mhi\  of $2 \times 10^6$ \Msol, the Smith Cloud has about the median \hi\ mass of the M33 HVCs 
and lies at the upper range 
of those around M31.   Given the observational uncertainties,  the Smith Cloud would be an inconspicuous addition to 
either galaxy \citep{Lockman2008,Westmeier2008,Keenan2016}.
However, as the brighter parts of the Smith Cloud lie only $\sim 3$ kpc away from the Galactic plane, 
if it were at the distance of M31 or M33 it would 
lie projected on the disk of those galaxies, and from even  the most favorable vantage would 
be separated by only $12^{\prime}$  from their disk.  It 
would almost certainly be considered part of the disk-halo interface and not cataloged as an HVC.

Some HVCs in M31 and M33 have linewidths $> 50$ \kms, suggesting large internal motions or disruption \citep{Westmeier2008,Keenan2016}. 
If the Smith Cloud were at the distance of M31, our poorer linear resolution would blend its different components, raising its \hi\ 
linewidth from the $<20$ \kms\ that we typically measure in its brighter parts, to  40 \kms.    

\section{\hi\ Outside Local Group Galaxies}
\label{sec:Interaction}

The Local Group contains two wonderful examples of \hi\ without stars, both associated with the Magellanic Clouds: 
 the Magellanic Stream (MS) and the Leading Arm (LA).  These are the rather enormous 
streams of gas (containing about 10\% the mass of the Milky Way's ISM) that are currently being 
lost from the Magellanic Clouds and are  
entering the Milky Way's CGM.  A distance of 55 kpc is assumed for both objects, though this has a large uncertainty; 
for a recent review see \cite{DOnghia2016}.    
The relevant 
\hi\  masses are $3 \times 10^8$ \Msol, and $3 \times 10^7$ \Msol, for the Stream and Leading Arm, respectively \citep{Bruns2005}.
The total mass of the Stream is $\sim 10^9$ \Msol, and is dominated by ionized gas \citep{Fox2014}.
The Magellanic Clouds are the only stellar systems in the Local Group that are observed to be losing their gas.  
The Magellanic Stream is important for our purposes because it  gives us a model 
for how we might interpret observations of other systems.  Although the 
combined MS + LA  is $200\arcdeg$\ long \citep{Putman2003,Nidever2010} 
 I estimate that no more than half this length would  appear with ${\rm N_{HI} > 10^{19}}$ \cms\ at
 a few Mpc distance, and thus be detectable in 21cm \hi\ emission 
in another galaxy group.    The detectable part of the MS 
 would have a length $\sim 100$ kpc and for the Leading Arm, about 60 kpc.  

Although we can see that the Stream and the Leading Arm are anchored in the stellar systems of the Magellanic Clouds, 
there are no stars associated with the gas of either object \citep{DOnghia2016}.  The absence of stars in the gas mirrors the fact that 
none of the stellar streams around the Milky Way or M31 have been found to have a convincing association with any neutral gas 
\citep{Lewis2013}.

The Milky Way's CGM seems quite resistant to the passage of neutral gas clouds even when they are attached to galaxies.  
Dwarf spheroidals within a few 100 kpc of the Milky Way seem utterly devoid of \hi:  the limits on some systems with 
$L_* \approx 2.5 \times 10^5$  L$_{\odot}$ are smaller than 100 M$_{\odot}$  \citep{Spekkens2014}.    A similar deficit is found 
for the dwarf satellites of M31 (Beaton, private communication).  
Because dwarf galaxies $>400$ kpc from the Milky Way seem to have retained their gas, 
the deficit for nearer dwarfs presumably reflects their stripping by the CGM  \citep{Blitz2000,Grcevich2009}.  
Distant dwarfs have ${\rm M_{HI}/L_{*}  \approx 1}$ \citep{Spekkens2014}, so we can 
estimate the \hi\ mass that has been deposited in the CGM through this stripping.
  For the dwarf galaxies currently known around the Milky Way, it amounts to ${\rm M_{HI} \approx 3\times 10^7}$ 
\Msol, an value almost identical to the estimated total \hi\ mass of HVCs in each of the HVC systems of the Milky Way, M31, and M33  
\citep{Putman2012,Westmeier2008,Keenan2016}.  It is tempting to dismiss  this 
as a coincidence, especially because the inferred mass of the ``stripped" \hi\ for the Milky Way is dominated by the most massive galaxy, 
the Sagittarius dSph \citep{Spekkens2014}.  But we do not have a good understanding of the 
inflow rate and location of the stripped material, which likely eventually make its way to the disk.

\section{Other Neutral Gas in the Local Group}
\subsection{IC 10}

\begin{figure}[b]
\includegraphics[width=4.5in]{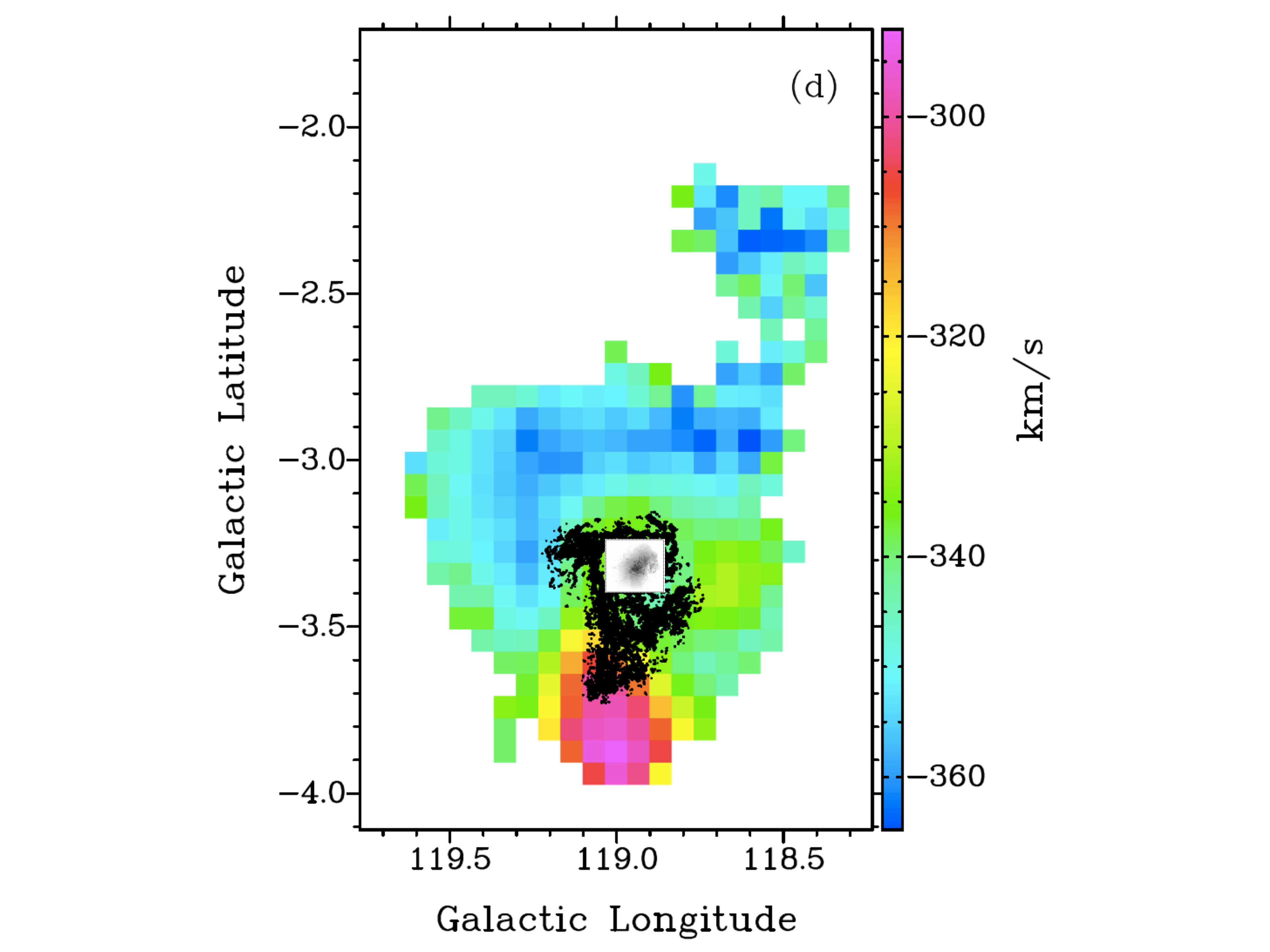}
\caption{Composite image of the stars and \hi\ of IC 10 adapted from \cite{Ashley2014}.  Colors show the \hi\ 
velocity field, highlighting the plume to the north.  Black shows \Nhi\ in the inner regions from 
 a high-resolution  map, showing the gas extension to the south.  The insert shows the stars of IC 10 in the optical V-band. 
}
\label{fig:IC10}       
\end{figure}

This  blue compact dwarf galaxy, a satellite of M31, is 
undergoing a burst of star formation and has a complex \hi\ structure with counter-rotating components and 
two tails or streamers shown in Figure \ref{fig:IC10} \citep{Shostak1989,Nidever2013,Ashley2014}.  
Kinematic analysis of the \hi\ suggests that the gas is inflowing rather than outflowing.  
Its stellar component, however, does not show 
evidence of streams or shells, or any kind of disturbance,  though the young stellar populations are spatially offset from 
the older stellar population  \citep{Gerbrandt2015}.   The data thus suggest that IC 10  is accreting nearly starless gas and not a companion.  
The mass in the southern and northern \hi\ features is $10^7$ \Msol\ and $6 \times 10^5$ \Msol, respectively, out of a total \mhi\ for the entire system of 
$8 \times 10^7$ \Msol.  The disturbed material is a significant fraction of the gas mass.  It is interesting that the two 
anomalous \hi\ features of IC 10 lie approximately on a line pointing back toward M31.

With a stellar mass of $8.6 \times 10^7$ \Msol\ \citep{McConnachie2012}, IC 10 has a ${\rm M_{HI}/L_V \approx 1}$, similar to 
gas-rich Local Group dwarfs, even though it is only 250 kpc from M31 and thus is not only a satellite of M31, 
but lies well within the CGM of that galaxy.    IC 10 and the Magellanic Clouds appear to be the only gas-rich dwarf galaxies 
within a few hundred kpc of the Milky Way or M31 in the Local Group, and the Magellanic Clouds are losing their gas while IC 10 seems 
to be accreting.

\subsection{M31-M33 Clouds}
\label{sec:M31-M33_clouds}
   
\begin{figure}[b]
\includegraphics[width=4.5in]{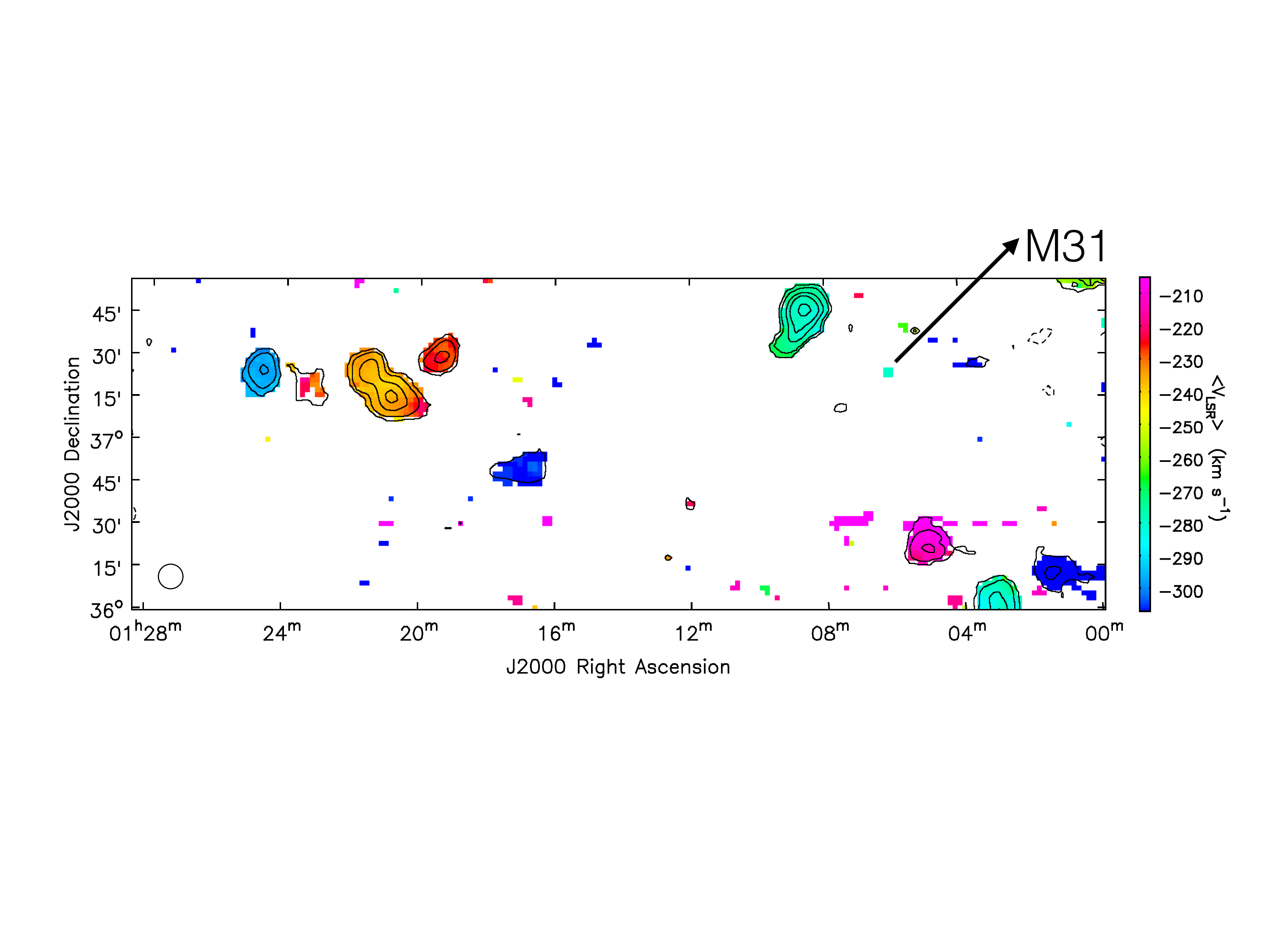}
\caption{Discrete \hi\ clouds found in a region about 100 kpc away from M31 in the 
direction of M33 \citep[adapted from][]{Wolfe2016}.  These are also displayed as the diamonds in Fig.~\ref{fig:VLGSR}.  
Contours mark  \Nhi\ in units of $5 \times 10^{17}$ \cms, scaled 
by -1 (dashed), 1, 2, 4, 6, 10.The typical cloud has ${\rm M_{HI} = 10^5}$ \Msol.  M31 lies to the upper right 
in the direction of the arrow; M33 to the lower left.  
}
\label{fig:M31M33_clouds}       
\end{figure}

In a map that reached a sensitivity to \Nhi\ of a few $10^{17}$ \cms, \cite{Braun_Thilker} detected very faint \hi\ emission 
in a long plume extending from M31 in the general direction of M33.  Subsequent observations with the GBT at higher 
angular resolution showed that most of the emission was concentrated into discrete clouds \citep{Wolfe2013,Wolfe2016}.   
Assuming that these clouds are at a distance of 800 kpc, between M31 and M33, they 
have ${\rm M_{HI} = 0.4-3.3 \times 10^5}$ \Msol, sizes of a few kpc, and are not associated with any stars or stellar stream.  
Figure~\ref{fig:M31M33_clouds} shows the clouds colored by their velocities.  The field lies $\sim 100$ kpc from M31.
  There are substantial differences in the 
velocities from cloud to cloud, $> 100$ \kms, indicating that these are discrete objects and not simply the brighter portions of a single extended sheet.
Velocity gradients across each cloud, in contrast, are only  $\approx 10$ \kms.  
The dynamical mass of the clouds, i.e., the mass needed for them to be self-gravitating, is typically  $10^3$ times more than the observed 
mass in \hi.  There are indications in the data that more 
clouds like these exist adjacent to the area covered by the GBT observations.

Fig.~\ref{fig:VLGSR} shows the relationship between the M31-M33 clouds and the galaxies M31 and M33, together 
with their systems of HVCs.  It is apparent that the M31-M33 clouds lack the large velocity spread of the HVCs, and that 
their kinematics has more in common with the systemic velocities of the galaxies than with their HVCs.  There is no 
apparent connection between these clouds and the M31 system of satellite galaxies \citep{Wolfe2016}.   

The M31-M33 clouds appear to be a new population in the Local Group with no known analogs.  
Beyond the work of \cite{Braun_Thilker} and the data shown in Fig.~\ref{fig:M31M33_clouds}, no other areas in 
the Local Group have been surveyed to the sensitivity necessary to detect \hi\ emission at these faint levels.  There is some 
indication that similar objects may exist to the north of M31 \citep{Wolfe2016} but current data are inconclusive.  \cite{Braun_Thilker} initially 
proposed that this \hi\ condensed from an intergalactic filament and there are models where it arises 
as the  result of an interaction between M31 and 
M33 \citep{Bekki2008,Lewis2013}.  The later suggestion, while attractive, now seems unlikely given the past history of the Local 
Group \citep{Shaya2013}.  We note that the M31-M33 clouds, like IC 10, reside well within the CGM of M31.  It is estimated that 
the CGM has a total column density ${\rm N_H \approx 10^{18}}$ \cms\ at the location of Fig.~\ref{fig:M31M33_clouds} \citep{Lehner2015}.  
The average \Nhi\  of the clouds over the entire field is only $9 \times 10^{16}$ \cms,  
so even if the ionization fraction of the CGM is $>90\%$, there would still be enough neutral material to account for the clouds \citep{Wolfe2016}.

\section{Starless \hi\ Near and Far}

This survey of \hi\ in the Local Group has ranged from the disk-halo clouds, which are certainly confined to the stellar disks, through the 
high velocity clouds, which seem preferentially located around the disks of galaxies, to material that is not 
incorporated into  galaxies: the Magellanic Stream and Leading Arm,  
the \hi\ streams intersecting IC 10, and finally the M31-M33 clouds.  
The first two, at least, are associated with stellar systems, while the M31-M33 clouds are not.  
Their properties are summarized in Table~\ref{tab:1}.  How common are 
objects like this in other galaxies and other groups?

\begin{figure}[b]
\includegraphics[width=4.5in]{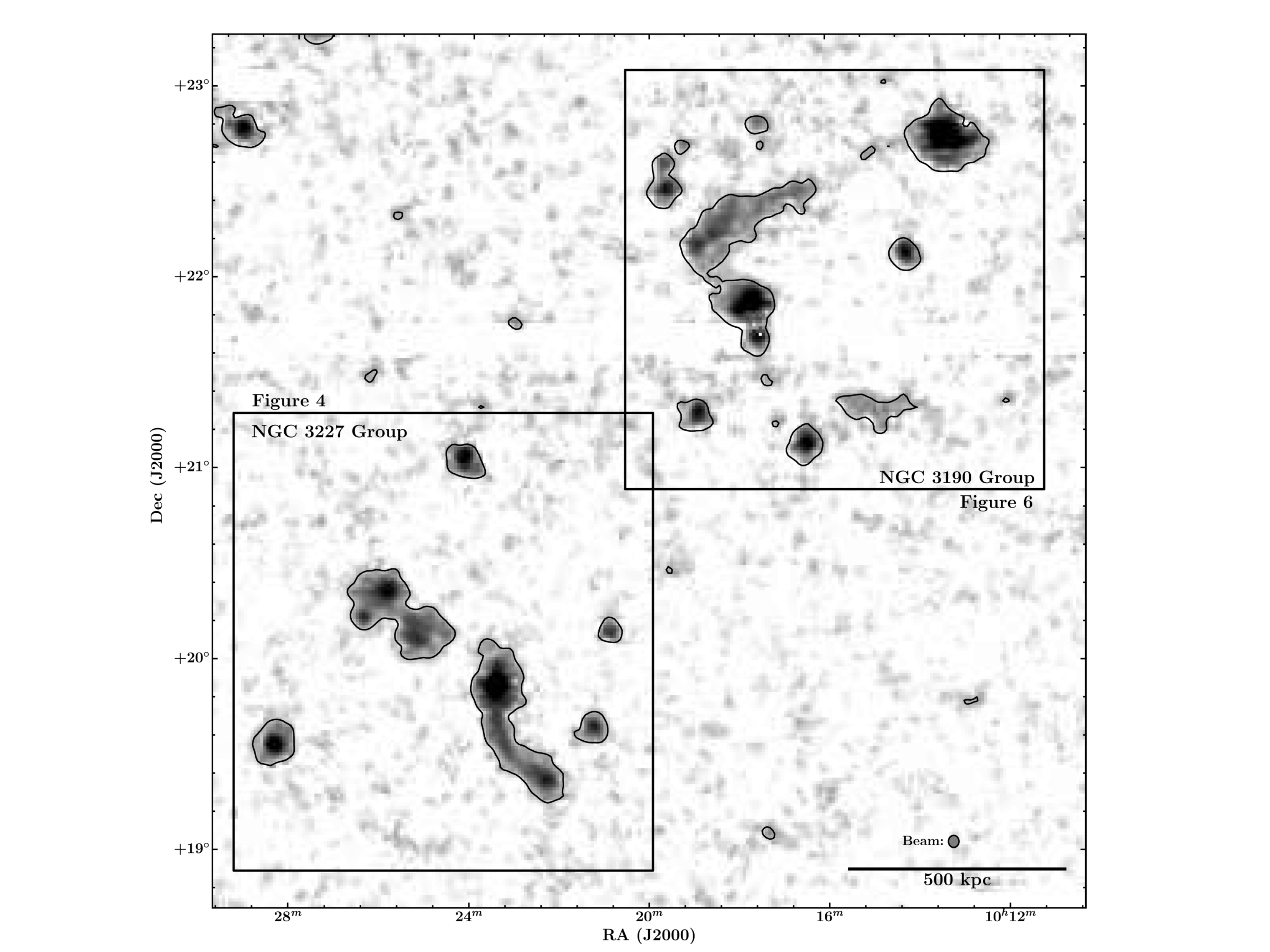}
\caption{Map of the \hi\  around the  NGC 3190 and NGC 3227 groups, showing extended tails and intragroup gas probably resulting from 
interactions between galaxies \citep[from][]{Leisman2016}.  
}
\label{fig:Leisman}       
\end{figure}

The disk-halo clouds and most HVCs would be blended together in current \hi\ observations of galaxies outside the Local Group, but 
similar material appears to be present in many other galaxies (see section \ref{sec:Disk-Halo}).    
In very sensitive \hi\  observations many, if not most spiral galaxies 
show \hi\ streams or features extending from the disk.  This 
 suggests that we are witnessing interaction, even though in many cases an accreting stellar companion cannot be identified \citep{Sancisi2008}.  True 
starless \hi\ clouds not associated with a galaxy seem rare, existing perhaps at the level of 2\% of the \hi\ objects detected 
in deep surveys, and even so, many of these seem to be located near large galaxies or be associated with tidal debris \citep{Kovac2009,Haynes2011}.  
Some of the most interesting objects, like the M31-M33 clouds and the plumes of IC 10, could be detected only out to a few 
 Mpc from the Milky Way with current instrumentation.

 The amount of \hi\ in starless gas in Local Group galaxies is rather modest compared to the enormous structures produced by 
 galaxy interactions in other groups.   The classic example is the  M81 group 
 \citep{Yun1994,Chynoweth2008}, but evidence for interactions like this in other galaxy groups abounds.  The data are too 
 numerous to summarize here, but as merely one example,  recent observations \citep{Leisman2016} have detected extended \hi\ features 600 kpc in length without stellar counterparts around the NGC 3590 and NGC 3227 groups  (Fig.~\ref{fig:Leisman}).

\begin{table}
\caption{Inventory of Starless \hi\ ``Clouds'' in the Local Group}
\label{tab:1}       
%
%
\begin{tabular}{p{4cm}p{2.1cm}p{1.6cm}p{2.8cm}p{1cm}}
\hline\noalign{\smallskip}
Object &  \mhi\ (\Msol) & Size (kpc) & Reference & Notes \\
\noalign{\smallskip}\svhline\noalign{\smallskip}
Disk-halo clouds & $7 \times 10^2$ & 0.06 & \cite{Ford2010}\\
Individual HVCs  & $\sim 10^6$       &  2-15  & \cite{Putman2012}\\
All HVCs  in MW, M31 or M33 & $2-5 \times 10^7$  & &  \cite{Westmeier2008,Putman2012,Keenan2016}\\
\hi\ stripped from MW dwarfs & $3 \times 10^7$  &   & \cite{Spekkens2014} \\
IC 10 northern stream & $6 \times 10^5$ & 7 & \cite{Nidever2013,Ashley2014}\\
IC 10 southern stream & $\sim 10^7$  & 7 & \cite{Nidever2013,Ashley2014}\\
Magellanic Stream & $3 \times 10^8$ & $\sim 100$ & \cite{Bruns2005,Nidever2010} &  a\\
Leading Arm          &  $3 \times 10^7$ & $\sim 60$ & \cite{Putman2003} & a \\
M31-M33 clouds & $\sim 10^5$ & 2  &  \cite{Wolfe2016} &  b\\
M31-M33 clouds all & $> 1.6 \times 10^6$ &  & \cite{Wolfe2016} & b\\
\noalign{\smallskip}\hline\noalign{\smallskip}
\end{tabular}
a.  Assumed distance 55 kpc.  b. Assumed distance 800 kpc.
\end{table}

If our Local Group is any example, the best place to detect starless neutral gas clouds is relatively close to galaxies.  As deep \hi\ 
surveys of the Local Group cover more area, we will obtain better insight into the origin of starless gas, and determine 
if it is always located near galaxies, or is more widespread.

\end{document}